\documentclass[aps,pre,twocolumn,10pt]{revtex4-1}

\usepackage{graphicx,amsmath,amssymb,bbm}
\usepackage[utf8x]{inputenc}

\DeclareMathOperator\erfc{erfc}

\begin{document}

\title{Transient superdiffusion in correlated diffusive media}

\author{Jacopo Bertolotti}
\affiliation{Physics and Astronomy Department, University of Exeter, Stocker 
Road, Exeter EX4 4QL, UK}

\begin{abstract}
Diffusion processes are studied theoretically for the case where the diffusion 
coefficient is itself a time and position dependent random function. We 
investigate how inhomogeneities and fluctuations of the diffusion coefficient 
affect the transport using a perturbative approach, with a special attention to 
the time scaling of the second moment. We show that correlated disorder can 
lead to anomalous transport and superdiffusion.
\end{abstract}

\maketitle

Despite its simplicity the diffusion equation is widely used to 
describe and model a large number of apparently unrelated systems. From 
heat~\cite{furierheat} and chemical~\cite{kramers} diffusion, to 
light in biological tissues~\cite{sheng, zaccanti} to electrons in impure 
metals~\cite{akkermansbook}, and even stock market 
fluctuations~\cite{bachelier, blackschole}. The underlying 
reason for its success is that many diverse systems can, at a microscopic 
level, be described via some form of random walk. Due to the Central Limit 
Theorem the diffusion equation is their appropriate macroscopic description 
irrespectively of the microscopic details~\cite{kolmogorv}. An emblematic 
characteristic of diffusive processes is the fact that the position variance 
grows linearly with time~\cite{rayleigh, einstein}. When the position variance 
scales either faster (superdiffusion) or slower (subdiffusion) than linear the 
transport is said to be anomalous~\cite{anotrans, klafterrestaurant}. Anomalous 
transport has been observed in the most diverse contexts: from transport in 
turbulent media~\cite{twoDturbolentLevy}, to earthquake 
patterns~\cite{Corral2006}, to light propagation in heterogeneous dielectric 
materials~\cite{WiersmaNature2008} and in hot atomic 
vapors~\cite{kaiserlevyatom}. In all these cases the hypothesis behind the 
Central Limit Theorem (and thus behind the diffusion approximation) are 
violated, either via a heavy tail in the step length distribution or a long 
memory kernel~\cite{rndwalkguide}. But once diffusion is established, it is 
assumed that no anomalous behavior is possible anymore. In fact, when the 
system is complex enough to appear random, it is often implicitly assumed that 
the fine structures of a position and/or time-dependent diffusion coefficient 
$D(\textbf{x},t)$ average out and that transport can be described via an 
effective diffusion constant $D_{\text{eff}}$, always leading to a linear 
scaling of the position variance.

Here we challenge this assumption. To do so we use a perturbative 
approach to study the effect on the position variance of a diffusion coefficient 
that fluctuates randomly in both position and time. In particular we show that 
the ensemble averaged transport can exhibit a transient superdiffusive behavior 
when $D(\textbf{x},t)$ fluctuations are correlated.

The diffusion equation can be obtained easily by coupling Fick's first 
law~\cite{fickslaw} with the continuity equation for a generic quantity $I$:
\begin{equation}
\label{eq:diff}
\left\{
 \begin{aligned}
  \mathbf{J} &= -D(\textbf{x},t) \nabla I \\
  \partial_t I&= -\nabla \cdot \mathbf{J}
 \end{aligned}
\right. \Rightarrow \partial_t I = \nabla \cdot D(\textbf{x},t) 
\nabla I ,
\end{equation}
where $\mathbf{J}$ is the flux of $I$. In the simplest case $D$ is a scalar 
constant and we retrieve the familiar form 
for the diffusion equation $\partial_t I=D \nabla^2 I$. Analytic solutions of 
Eq.~\ref{eq:diff} are known only for simple geometries, e.g. a stationary 
multilayer system~\cite{multilayerdiff}.
\begin{figure}[tb]
	\centering
		
\includegraphics[width=0.4\textwidth]{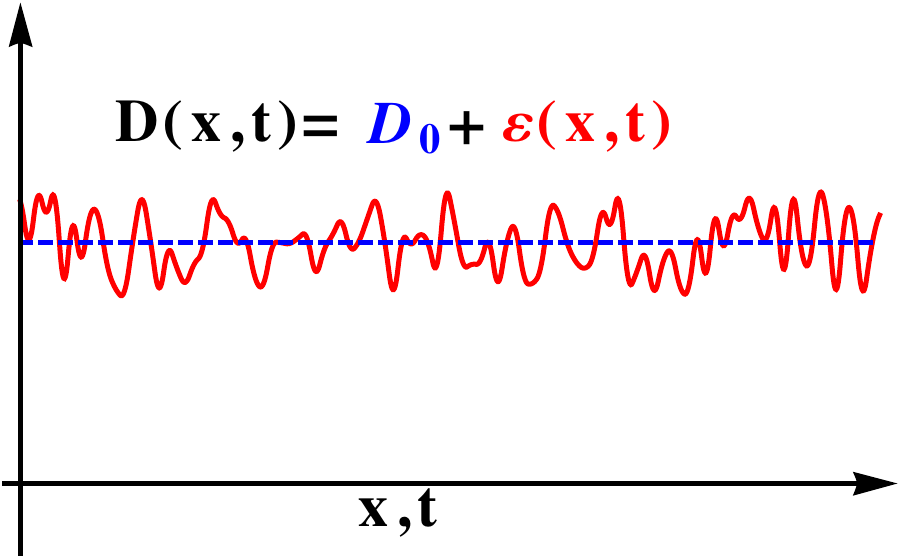}
	\caption{The time and position-dependent diffusion coefficient 
$D(\textbf{x},t)$ can be decomposed in a constant part $D_0$ (blue dashed line) 
and a fluctuating part $\varepsilon (\textbf{x},t)$ (red continuous line).}
	\label{fig:D_decomposition}
\end{figure}
Instead of looking for a general solution, we decompose the diffusion 
coefficient in a constant part $D_0$ and a fluctuating part 
$\varepsilon(\textbf{x},t)$ as shown in Fig.~\ref{fig:D_decomposition}, 
rewriting Eq.~\ref{eq:diff} as $\partial_t I = D_0 \nabla^2 I + \nabla 
\varepsilon (\textbf{x},t) \nabla I$. If both $\varepsilon$ and its gradient 
are small enough, we can employ a perturbative approach to yield
\begin{equation}
\label{eq:perturbation}
 I = g \odot \left[ S +  \nabla \cdot \varepsilon (\textbf{x},t) \nabla I 
\right]
\end{equation}
where $S(\mathbf{x},t)$ is the source term, $\odot$ represent the convolution 
product with respect of both the spatial and temporal coordinates defined in 
$d$ dimensions as
\begin{equation*}
\label{eq:odot}
 f \odot h = \int_{\mathbbm{R}^d} d\mathbf{y} \int_0^t d \tau f \left( 
\mathbf{x}- \mathbf{y}, t - \tau \right) h \left( \mathbf{y} , \tau \right) ,
\end{equation*}
and $g$ is the Green's function associated to the unperturbed diffusion 
equation
\begin{equation*}
 g \left( \mathbf{x}, t \right) = \frac{\Theta \left( t \right) }{\sqrt{\left( 
4 \pi D_0 t \right)^d}} 
e^{-\frac{\mathbf{x}^2}{4 D_0 t}},
\end{equation*}
where $\Theta$ is the Heaviside step function. We notice that $g$ is such 
that $f \odot g = g \odot f$ for every $f$ such that $f \odot g$ exists. For 
simplicity we will consider only sources terms of the form $S= 
\delta(\mathbf{x}) \delta(t)$. More complicated sources can easily be 
incorporated, but do not change the general results.

Iterating Eq.~\ref{eq:perturbation} we obtain the perturbative series $I= I_0 + 
I_1 + I_2 +\dots$ where
\begin{equation*}
 \begin{aligned}
  I_0 &= g \odot S = g \\
  I_1 &= g \odot \left[ \nabla \cdot \left( \varepsilon \nabla \left( g \odot S 
\right) \right) \right] \\
  &= g \odot \left[ \nabla \cdot \left( \varepsilon \nabla g \right) \right] \\
  I_2 &=  g \odot \left[ \nabla \cdot \left( \varepsilon \nabla \left( g \odot 
\left( \nabla \cdot \left( \varepsilon \nabla \left( g \odot S \right) \right)  
\right) \right) \right) \right] \\
  &= g \odot \left[ \nabla \cdot \left( \varepsilon \nabla \left( g \odot 
\left( \nabla \cdot \left( \varepsilon \nabla g \right)  \right) \right) \right) 
\right] \\
 I_3 &= \dots
 \end{aligned}
\end{equation*}
and we used the fact that $g \odot S = g$.

In order to characterize a possible anomalous scaling of the position variance, 
we define  the $n$th moment along the $x_i$ direction of a generic function 
$f$ as
\begin{equation*}
 \mu^{(n)}_{x_i} \left[ f \right] = \int_{-\infty}^{+\infty} \left( x_i 
\right)^n f \left(\mathbf{x},t \right) d\mathbf{x}
\end{equation*}
that satisfies these useful properties~\cite{mellin,convolutionmoments}:
\begin{equation}
\label{eq:prop1}
 \mu^{(n)}_{x_i} \left[ \partial_j f \right] = \left\{
\begin{aligned}
-& n \mu^{(n-1)}_{x_j} \left[f \right] \quad & i=j\\
0 & & i\neq j
\end{aligned}
 \right.
\end{equation}
and
\begin{equation}
\label{eq:prop2}
\begin{aligned}
\mu^{(0)} \left[ f \odot g \right] &= \mu^{(0)} \left[ f \right] \circledast 
\mu^{(0)} \left[ g \right] ,\\
 \mu^{(1)}_{x_i} \left[ f \odot g \right] &= \mu^{(1)}_{x_i} \left[ f \right] 
\circledast \mu^{(0)} \left[g \right] + \mu^{(0)} \left[ f \right] 
\circledast \mu^{(1)}_{x_i} \left[ g \right] ,\\
\mu^{(2)}_{x_i} \left[ f \odot g \right] &= \mu^{(2)}_{x_i} \left[ f \right] 
\circledast \mu^{(0)} \left[g \right] + \mu^{(0)} \left[ f \right] 
\circledast \mu^{(2)}_{x_i} \left[ g \right] +\\
+2 & \mu^{(1)}_{x_i} \left[ f \right] 
\circledast \mu^{(1)}_{x_i} \left[ g \right]
\end{aligned}
\end{equation}
where $\circledast$ represents a convolution product with respect to the sole 
time variable.

Using these properties we can calculate the zeroth moment for 
every term in the perturbation:
\begin{equation*}
 \mu^{(0)} \left[ I_0 \right] =  \mu^{(0)} \left[g \right] = 1
\end{equation*}
where we used the fact that the Green's function $g$ is normalized to 1. 
For the first perturbative order we get
\begin{equation*}
\begin{aligned}
\mu^{(0)} & \left[ I_1 \right] =  \mu^{(0)} \left[  g \odot \left( \nabla 
\cdot \left( \varepsilon \nabla g \right) \right) \right] = \sum_{i=1}^d 
\mu^{(0)} \left[  g \odot \partial_i \left( \varepsilon \partial_i g  \right) 
\right] \\
&= \sum_{i=1}^d \mu^{(0)} \left[ \partial_i g \odot \left( \varepsilon 
\partial_i g  \right) \right] = \sum_{i=1}^d \mu^{(0)} \left[ \partial_i g 
\right] \circledast \mu^{(0)} \left[\varepsilon \partial_i g  \right] \\ &= 0
\end{aligned}
\end{equation*}
where we used the fact that $g$ is an even function and thus $\mu^{(0)} \left[ 
\partial_i g \right]=0$. Since the leftmost $g$ and the leftmost $\nabla$ 
produce a term proportional to $\mu^{(0)} \left[ \partial_i g \right]$ for 
every perturbative order, we see that the zeroth moment is non null only for 
the unperturbed term $I_0$. Therefore the perturbative series is correctly 
normalized at all orders.

Similarly for the first moment we obtain:
\begin{equation*}
 \mu^{(1)}_{x_i} \left[ I_0 \right] = \mu^{(1)}_{x_i} [g] =0 ,
\end{equation*}
\begin{equation*}
 \begin{aligned}
  \mu^{(1)}_{x_i} \left[ I_1 \right] &= \sum_{j=1}^d \mu^{(1)}_{x_i} \left[ 
\partial_j g \odot \left( \varepsilon \partial_j g \right) \right] \\
&= \sum_{j=1}^d \mu^{(1)}_{x_i} \left[ \partial_j g \right] \circledast 
\mu^{(0)} \left[  \varepsilon \partial_j g \right] .
 \end{aligned}
\end{equation*}
Performing an ensemble average $\left\langle \cdot \right\rangle$ over all 
possible realization of $\varepsilon$ we obtain
\begin{equation*}
 \left\langle \mu^{(1)}_{x_i} \left[ I_1 \right] \right\rangle = \left\langle 
\varepsilon \right\rangle \sum_{j=1}^d \mu^{(1)}_{x_i} \left[ \partial_j g 
\right] \circledast 
\mu^{(0)} \left[ \partial_j g \right] =0 .
\end{equation*}
For the second perturbative term we get:
\begin{equation*}
 \begin{aligned}
  \left\langle \mu^{(1)}_{x_i} \left[ I_2 \right] \right\rangle &=\\
&= \left\langle \sum_{j,k=1}^d \mu^{(1)}_{x_i} \left[ \partial_j g \odot \left( 
\varepsilon \left( \partial_{j} \partial_k g \odot \left( \varepsilon 
\partial_k g \right) \right) \right) \right] \right\rangle \\
&= - \left\langle \sum_{i=1}^d 1 \circledast \mu^{(0)} \left[ \varepsilon \left(
\partial_i^2 g \odot \varepsilon \partial_i g \right) \right] \right\rangle \\
&= - \sum_{i=1}^d 1 \circledast \mu^{(0)} \left[ \left( \left\langle 
\varepsilon 
\varepsilon \right\rangle \partial_{i}^2 g \right) \odot \left( \partial_i g 
\right) \right]\\
&= - \sum_{i=1}^d 1 \circledast \mu^{(0)} \left[ \left\langle \varepsilon 
\varepsilon \right\rangle \partial_{i}^2 g \right] \circledast \mu^{(0)} \left[ 
\partial_i g \right] =0 ,
\end{aligned}
\end{equation*}
where we used the fact that, on average, the system is translational invariant 
in both space and time, and thus $\left\langle \varepsilon (\mathbf{x},t) 
\varepsilon (\mathbf{x}^{\prime},t^{\prime}) \right\rangle = \left\langle 
\varepsilon \varepsilon \right\rangle \left( | \mathbf{x} - \mathbf{x}^{\prime}| 
, | t - t^{\prime} | \right)$. We can see that, similarly to what happened for 
the zeroth moments, the last term in the ensemble averaged first moment can 
always be written as $\mu^{(0)} \left[ \partial_i g \right]$. As a consequence, 
correlations in the disorder at any perturbative order do not change the centre 
of mass position of $I$.

The calculation of the second moment is more delicate and requires some 
attention. For the unperturbed term the second moment is trivially
\begin{equation*}
 \mu^{(2)}_{x_i} \left[ I_0 \right] = \mu^{(2)}_{x_i} [g] = 2 D_0 t .
\end{equation*}

For the first perturbative term we get:
\begin{equation*}
 \begin{aligned}
 \left\langle \mu^{(2)}_{x_i} \left[ I_1 \right] \right\rangle &= 
\left\langle \sum_{j=1}^d \mu^{(2)}_{x_i} \left[ \partial_j g \odot \left( 
\varepsilon \partial_j g \right) \right] \right\rangle \\
&= 2 \left\langle \sum_{j=1}^d \mu^{(1)}_{x_i} \left[ \partial_j g \right] 
\circledast \mu^{(1)}_{x_i} \left[ \varepsilon \partial_j g \right] 
\right\rangle \\
&= -2 \left( 1 \circledast \mu^{(1)}_{x_i} \left[ \left\langle \varepsilon 
\right\rangle \partial_i g 
\right] \right) \\
&= -2 \left\langle \varepsilon \right\rangle \left( 1 \circledast 
\mu^{(1)}_{x_i} \left[ \partial_i g \right] \right) = 2 \left\langle \varepsilon 
\right\rangle t .
\end{aligned}
\end{equation*}
Therefore the average effect of a inhomogeneous and fluctuating diffusion 
coefficient, up to the first perturbative order, can be captured by using an 
effective diffusion constant $D_{\text{eff}}=D_0 + \left\langle \varepsilon 
\right\rangle$. We can thus safely assume in the following that $\left\langle 
\varepsilon \right\rangle =0$.

All the averaged moments of the second perturbative order $I_2$ depend 
explicitly on the two-point correlation $\left\langle \varepsilon \varepsilon 
\right\rangle$, but the second moment is the first one that is not identically 
zero:
\begin{equation}
\label{eq:mu2i2}
 \begin{aligned}
 \left\langle \mu^{(2)}_{x_i} \left[ I_2 \right] \right\rangle &=\\
&= \left\langle \sum_{j,k=1}^d \mu^{(2)}_{x_i} \left[ \partial_j g \odot \left( 
\varepsilon \left( \partial_{j} \partial_k g \odot \varepsilon \left( 
\partial_k g \right) \right) \right) \right] \right\rangle \\
&= 2 \left\langle \sum_{j,k=1}^d \mu^{(1)}_{x_i} \left[ \partial_j g \right] 
\circledast \mu^{(1)}_{x_i} \left[ \varepsilon \left( \partial_{j} \partial_k g 
\odot \varepsilon \partial_k g \right) \right] \right\rangle \\
&= -2 \left\langle \sum_{k=1}^d 1 \circledast \mu^{(1)}_{x_i} \left[ 
\varepsilon \left( \partial_{i} \partial_k g \odot \varepsilon \partial_k g 
\right) \right] \right\rangle \\
&= -2 \sum_{k=1}^d 1 \circledast \mu^{(1)}_{x_i} \left[ \left( \left\langle 
\varepsilon \varepsilon \right\rangle \partial_{i} \partial_k g \right) \odot 
\left( 
\partial_k g \right) \right] \\
&= 2 \left( 1 \circledast 1 \circledast \mu^{(0)} \left[ \left\langle 
\varepsilon \varepsilon \right\rangle \partial_{i}^2 g \right] \right) \\
&= 2 \int_0^t \left( t-\tau \right) \left( \int_{\mathbbm{R}^d} 
\left\langle 
\varepsilon \varepsilon \right\rangle \partial_i^2 g \left( \mathbf{x}, \tau 
\right) \, d\mathbf{x} \right) d\tau
 \end{aligned}
\end{equation}
where in the last step we used the Cauchy formula for repeated integration.

Eq.~\ref{eq:mu2i2} shows that the presence of correlations can have an 
influence on the scaling properties of the second moment. To study the nature 
and extension of this influence we focus on the 3D case ($d=3$) and assume that 
the system is not only translational invariant but also isotropic. 
Eq.~\ref{eq:mu2i2} can thus be naturally rewritten in spherical coordinates as
\begin{equation}
\label{eq:mu2i2r}
 \mu^{(2)}_{r} \left[ \left\langle I_2 \right\rangle \right] = \frac{8 \pi}{3} 
\int_0^t 
\left( t-\tau \right) \left( \int_{0}^{\infty} 
\left\langle 
\varepsilon \varepsilon \right\rangle r^2 \nabla^2_r g \, dr 
\right) d\tau ,
\end{equation}
where $\nabla^2_r = r^{-2} \partial_r r^2 \partial_r$ is the radial part 
of the Laplacian operator in spherical coordinates.
\begin{figure}[htb]
	\centering
\includegraphics[width=0.4\textwidth]{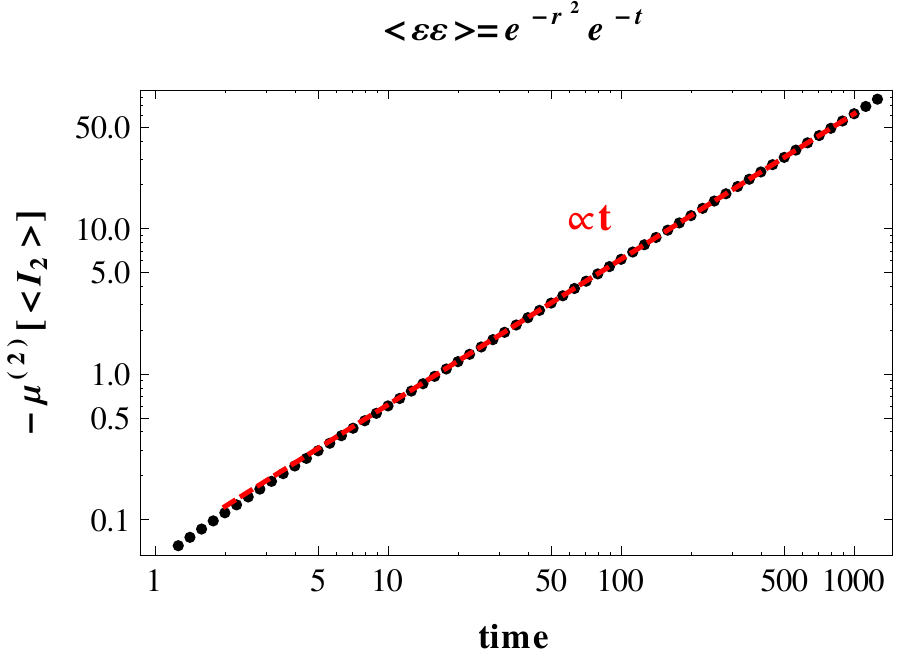}
	\caption{Numerical solution of Eq.~\ref{eq:mu2i2r} for $D_0=1$ and 
$\left\langle \varepsilon \varepsilon \right\rangle = e^{-r^2} e^{-t}$ (black 
dots). Short range (but not vanishing) correlations lead to a contribution to 
the second moment that scale linearly with time (red dashed line), and 
thus can be absorbed into an effective diffusion constant.}
	\label{fig:short_corr}
\end{figure}
\begin{figure}[tb]
	\centering
\includegraphics[width=0.4\textwidth]{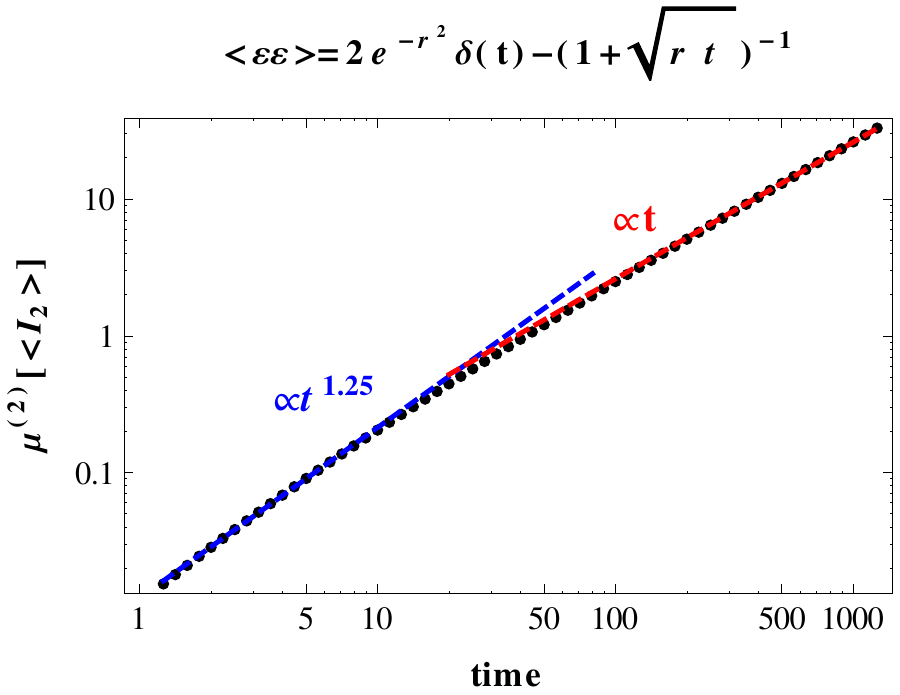}
	\caption{Numerical solution of Eq.~\ref{eq:mu2i2r} for $D_0=1$ and 
$\left\langle \varepsilon \varepsilon \right\rangle = 2 e^{-r^2} \delta (t) - 
\left( 1+ \sqrt{r t} \right)^{-1}$ (black dots). The short range term $2 
e^{-r^2} \delta (t)$ give us the correct limit of the correlations when both $r$ 
and $t$ go to zero, but does not influence $\mu^{(2)}_{x_i} \left[ \left\langle 
I_2 \right\rangle \right]$. The long range anticorrelation term leads to a 
transient superdiffusive term that scales as $t^{1.25}$ (blue dashed line).}
	\label{fig:long_corr}
\end{figure}
\begin{figure}[tb]
	\centering
\includegraphics[width=0.4\textwidth]{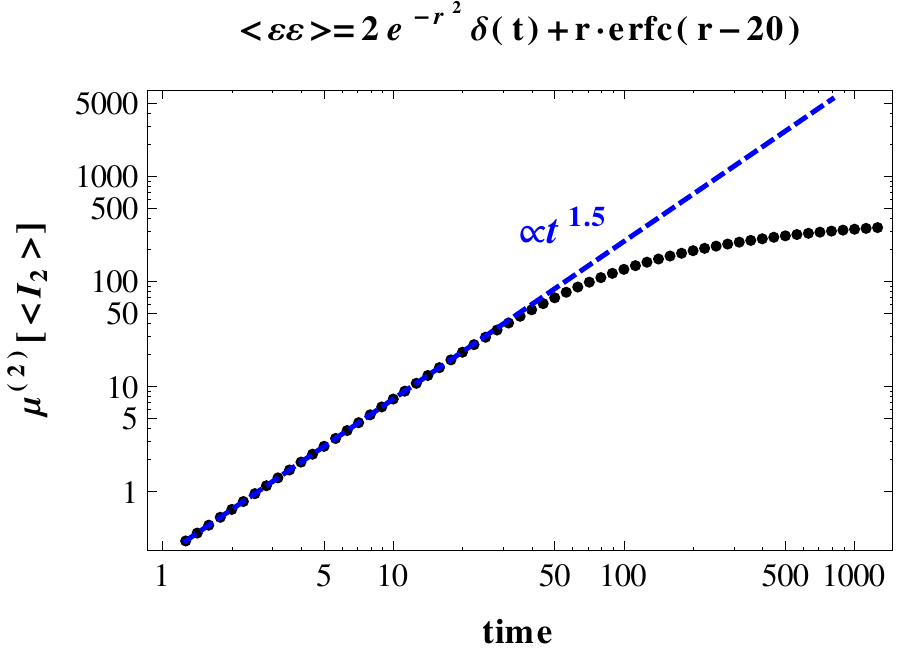}
	\caption{Numerical solution of Eq.~\ref{eq:mu2i2r} for $D_0=1$ and 
$\left\langle \varepsilon \varepsilon \right\rangle = 2 e^{-r^2} \delta (t)+r 
\erfc (r-20)$ (black dots), where $\erfc$ is the complementary error function 
that acts as a truncation for the correlations. As in Fig.~\ref{fig:long_corr} 
the short range term is needed to recover the correct limit for $r=t=0$. The 
rising (but truncated) time-independent correlations leads to a transient 
superdiffusive term that scales as $t^{1.5}$ (blue dashed line). At long times 
this contribute saturates to a finite value.}
	\label{fig:rising_corr}
\end{figure}
In general we can write the 2-point correlation as $\left\langle \varepsilon 
\varepsilon \right\rangle = \left\langle \varepsilon^2 \right\rangle h \left( 
r, \tau \right)$ where $\left\langle \varepsilon^2 \right\rangle$ represents 
the variance of the diffusion coefficient's fluctuations (that we will set to 
1 unless explicitly stated), and $h \left( r, \tau \right)$ is a distribution 
that describe the shape of the correlations. In order to be physical we 
require $h$ to go to zero when either $r$ or $\tau$ go to infinity, and to go 
to 1 (or to a Dirac delta) when both $r$ and $\tau$ go to zero. If the time 
fluctuations of $\varepsilon$ are well represented by white noise, then 
$\left\langle \varepsilon \varepsilon \right\rangle \propto \delta (\tau)$. 
Since $\lim_{t\rightarrow 0} \nabla^2 g=0$ there will be no contribution to the 
second moment. More complicated functional forms for the two-point 
correlations require a numerical integration of Eq.~\ref{eq:mu2i2r}.

The effect of a given  form for the correlations can be categorized in a 
limited numbers of cases.
\begin{enumerate}
 \item It can produce a null contribution.
 \item It can produce a contribution that scale linearly with time (either 
positive or negative). In this case the effect can be absorbed in the effective 
diffusion constant $D_{\text{eff}}$.
\item It can produce a contribution that grows slower than linearly. In this 
case the contribution from the unperturbed term is always bigger than the one 
due to the correlations, and thus it is negligible.
\item It can produce a contribution that grows faster than linearly. In this 
case the overall transport is effectively superdiffusive.
\end{enumerate}
Not surprisingly we find that short range correlations always yield one of the 
first 3 cases. A typical example is shown in Fig.~\ref{fig:short_corr}.

For long range correlations of the form $\left\langle \varepsilon \varepsilon 
\right\rangle \propto r^a t^b$, Eq.~\ref{eq:mu2i2r} can be integrated to give 
$\mu^{(2)}_{x_i} \left[ \left\langle I_2 \right\rangle \right] \propto 
t^{1+b+a/2}$. If at least one of $a$ or $b$ is positive then the transport can 
be superdiffusive. It is interesting to notice that if we allow $\left\langle 
\varepsilon^2 \right\rangle$ to grow polynomially with time, the position 
variance can grow as fast as we want, even faster than the ballistic $t^2$ 
case. Of course increasing the fluctuations requires a steady influx of energy 
into the system and therefore an accelerated transport regime is not an 
impossibility. Furthermore, sooner or later the growing fluctuations will 
violate the assumptions behind our perturbative approach and thus this result 
can not be extrapolated to the large time limit. Since $\left\langle \varepsilon 
\varepsilon \right\rangle$ can not grow indefinitively, neither $a$ or 
$b$ can be positive at large time or large distances. This results in a 
diffusive transport in the long time limit. This does not mean that there can 
not be a transient superdiffusive regime similar to the one encountered in 
truncated Lévy walks~\cite{mantegna}. Fig~\ref{fig:long_corr} shows the 
transient superdiffusive transport where the second moment grows as $t^{1.25}$ 
due to a (anti)correlation that decays asymptotically as $(\sqrt{r t})^{-1}$. 
Finally, while it is true that the correlation function can not grow 
indefinitely, if we truncate it after a certain distance/time we obtain again a 
transient superdiffusive transport, as shown in Fig.~\ref{fig:rising_corr}.

In conclusion we showed that correlations in the diffusion coefficient 
fluctuations can lead to a transient superdiffusive behavior, and found an 
explicit formula to link the two-point correlation $\left\langle \varepsilon 
\varepsilon \right\rangle$ with the time scaling of the position variance. The 
higher order perturbation terms depend on the three-point correlation 
$\left\langle \varepsilon \varepsilon \varepsilon \right\rangle$, four-point 
correlation $\left\langle \varepsilon \varepsilon \varepsilon \varepsilon 
\right\rangle$ and so on, that can also lead to deviations from a standard 
diffusive transport.

\begin{acknowledgments}
We thank Janet Anders, Simon Horsley and Thomas Philbin for helpful discussion 
and invaluable suggestions.
\end{acknowledgments}


\begin{thebibliography}{99}
\bibitem{furierheat}
J. Fourier, \textit{Théorie analytique de la chaleur} (Firmin Didot Père et 
Fils, Paris, 1822).
\bibitem{kramers}
H.A. Kramers, Physica \textbf{7}, 284 (1940).
\bibitem{sheng}
P. Sheng, \textit{Introduction to Wave Scattering, Localization and Mesoscopic 
Phenomena} (Springer, 2010).
\bibitem{zaccanti}
F. Martelli, S. Del Bianco, A, Ismaelli, G. Zaccanti, \textit{Light Propagation 
Through Biological Tissue and Other Diffusive Media} (SPIE Press, 2009).
\bibitem{akkermansbook}
E. Akkermans and G. Montambaux, \textit{Mesoscopic Physics of Electrons and
Photons} (Cambridge University Press, 2007).
\bibitem{bachelier}
L. J.-B. A. Bachelier, Ann. Sci. Ec. Norm. Sup. \textbf{3} 21 (1900).
\bibitem{blackschole}
F. Black and M. Scholes, J. Polit. Econ. \textbf{81}, 637 (1973).
\bibitem{kolmogorv}
B. V. Gnedenko and A. N. Kolmogorov, \textit{Limit Distributions for Sums of 
Independent Random Variables} (Addison-Wesley, 1954).
\bibitem{rayleigh}
K. Pearson and J. W. Strutt, Nature \textbf{72}, 294; 318; 342 (1905).
\bibitem{einstein}
A. Einstein, \textit{Investigations on the Theory of the Brownian Movement} 
edited by R. Furth (Dover Publications, New York, 1998).
\bibitem{anotrans}
R. Klages, G. Radons, and I. M. Sokolov, \textit{Anomalous Transport}
(Wiley-VCH, 2008); R. Metzler, and J. Klafter, Phys. Rep. \textbf{339}, 1
(2000).
\bibitem{klafterrestaurant}
R. Metzler and J. Klafter, J. Phys. A: Math. Gen. \textbf{37}, R161 (2004).
\bibitem{twoDturbolentLevy}
T. H. Solomon, E. R. Weeks, and H. L. Swinney, Phys. Rev. Lett. \textbf{71}, 
3975 (1993).
\bibitem{Corral2006}
A. Corral, Phys. Rev. Lett. \textbf{97}, 178501 (2006).
\bibitem{WiersmaNature2008}
P. Barthelemy, J. Bertolotti, and D. S. Wiersma, Nature \textbf{453}, 495
(2008).
\bibitem{kaiserlevyatom}
N. Mercadier, W. Guerin, M. Chevrollier, and R. Kaiser, Nature Phys. \textbf{5},
602 (2009).
\bibitem{rndwalkguide}
R. Metzler and J. Klafter, Phys. Rep. \textbf{339}, 1 (2000).
\bibitem{fickslaw}
A. Fick, Ann. der. Physik \textbf{94}, 59 (1855).
\bibitem{multilayerdiff}
R. I. Hickson, S. I. Barry, G. N. Mercer, Int. J. Heat Mass Tran. \textbf{52}, 
5776 (2009).
\bibitem{mellin}
A. D. Poularikas, \textit{Handbook of Formulas and Tables for Signal 
Processing} (Springer, 1999).
\bibitem{convolutionmoments}
C. A. Laury-Micoulaut, Astron. Astrophys. \textbf{51} 343 (1976).
\bibitem{mantegna}
R. N. Mantegna and H. E. Stanley, Phys. Rev. Lett. \textbf{73}, 2946 (1994).
\end{thebibliography}
\end{document}